\documentclass[a4paper,11pt]{article}
\pdfoutput=1 
\usepackage{jcappub} 
\usepackage{amsmath,amsfonts,amssymb,mathtools}
\usepackage[numbers]{natbib}
\usepackage[T1]{fontenc} 
\usepackage{booktabs}
\usepackage{multirow}
\usepackage{xcolor}
\usepackage{graphicx}
\usepackage{hyperref}
\hypersetup{colorlinks,allcolors=blue}
\usepackage{subcaption}
\newcommand{\effort}{\texttt{Effort.jl}}
\newcommand{\julia}{\texttt{Julia}}
\newcommand{\scs}{\texttt{SimpleChains.jl}}
\newcommand{\lux}{\texttt{Lux.jl}}
\newcommand{\capse}{\texttt{Capse.jl}}
\newcommand{\turing}{\texttt{Turing.jl}}
\newcommand{\camb}{\texttt{CAMB}}
\newcommand{\class}{\texttt{CLASS}}

\newcommand{\matryoshka}{\texttt{Matryoshka}}
\newcommand{\emulatelss}{\texttt{EmulateLSS}}
\newcommand{\comet}{\texttt{Comet}}
\newcommand{\montepython}{\texttt{MontePython}}

\newcommand{\cosmopower}{\texttt{CosmoPower}}
\newcommand{\odejl}{\texttt{DifferentialEquations.jl}}
\newcommand{\tullio}{\texttt{Tullio.jl}}

\newcommand{\pybird}{\texttt{pybird}}
\newcommand{\blast}{\texttt{Blast.jl}}
\usepackage{aas_macros}

\newcommand{\ace}{\texttt{AbstractCosmologicalEmulators.jl}}

\title{\effort{}: a fast and differentiable emulator for the Effective Field Theory of the Large Scale Structure of the Universe}
\author[a,b,c,1]{Marco Bonici,\note{Corresponding author}}
\author[d,e]{Guido D'Amico,}
\author[f]{Julien Bel,}
\author[c]{ and Carmelita Carbone}

\affiliation[a]{Waterloo Centre for Astrophysics, University of Waterloo, Waterloo, ON N2L 3G1, Canada}
\affiliation[b]{Department of Physics and Astronomy, University of Waterloo, 200 University Ave W, Waterloo, ON N2L 3G1, Canada}
\affiliation[c]{INAF Istituto di Astrofisica Spaziale e Fisica cosmica di Milano, Via Alfonso Corti 12, I-20133 Milano, Italy}
\affiliation[d]{Dipartimento di Scienze Matematiche, Fisiche e Informatiche, Università di Parma, Viale delle Scienze 7/A 43124 Parma, Italy}
\affiliation[e]{INFN Gruppo Collegato di Parma, Viale delle Scienze 7/A 43124
Parma, Italy}
\affiliation[f]{Aix Marseille Univ, Université de Toulon, CNRS, CPT, Marseille, France}
\date{\today}

\emailAdd{mbonici@uwaterloo.ca}

\begin{document}

\abstract{
We present the official release of the EFfective Field theORy surrogaTe (\effort{}), a novel and efficient emulator designed for the Effective Field Theory of Large-Scale Structure (EFTofLSS). This tool combines state-of-the-art numerical methods and clever preprocessing strategies to achieve exceptional computational performance without sacrificing accuracy. To validate the emulator reliability, we compare Bayesian posteriors sampled using \effort{} via Hamiltonian MonteCarlo methods to the ones sampled using the widely-used \pybird{} code, via the Metropolis-Hastings sampler. On a large-volume set of simulations, and on the BOSS dataset, the comparison confirms excellent agreement, with deviations compatible with MonteCarlo noise. Looking ahead, \effort{} is poised to analyze next-generation cosmological datasets and to support joint analyses with complementary tools.}


\maketitle
\flushbottom

\section{Introduction} 
\label{sec:introduction}
In the era of precision cosmology, the analysis of Large Scale Structure (LSS) surveys play a pivotal role in understanding the universe's contents and dynamics, from its birth to late-time accelerated expansion. Forthcoming datasets from collaborations such as the Dark Energy Spectroscopic Instrument (DESI)~\cite{DESI:2016fyo, DESI:2022xcl, DESI:2024hhd} and Euclid~\cite{Euclid:2024yrr} promise to probe the universe’s LSS with unprecedented accuracy. These datasets present opportunities to explore both the standard cosmological model and its potential extensions with unparalleled precision. However, they also impose new computational demands because of the increased dimensionality of the posterior distribution to explore, requiring tools that can efficiently handle the scale and complexity of these analyses.

One promising technique that has emerged to address these computational challenges is emulation. An emulator is a surrogate model that can approximate the outputs of computationally expensive models, but at a significantly lower computational cost. This approach offers a twofold benefit: it accelerates the computation and allows for the application of more advanced optimization and sampling techniques, in particular gradient-based methods. In recent years, emulators have proven to be highly successful in domains with computationally demanding theoretical models, such as in cosmology.

After the first seminal papers on this topic~\cite{jimenez_fast_2004, fendt_pico_2007, auld_fast_2007}, several emulators have been produced in the literature, emulating the output of Boltzmann solvers such as \camb{}~\cite{lewis_efficient_2000} or \class{}~\cite{blas_cosmic_2011}, with applications ranging from the Cosmic Microwave Background (CMB)~\cite{albers_cosmicnet_2019,  SpurioMancini:2021ppk, nygaard_connect_2022, gunther_cosmicnet_2022, capse}, the linear matter power spectrum~\citep{Mootoovaloo:2021rot, SpurioMancini:2021ppk, Donald-McCann:2021nxc, arico_accelerating_2021, Bartlett:2023cyr, Bakx:2024zgu}, galaxy power spectrum multipoles~\citep{arico_accelerating_2021, Eggemeier:2022anw, Donald-McCann:2022pac, Trusov:2024mmw, Bakx:2024zgu}, and the galaxy survey angular power spectrum~\citep{manrique-yus_euclid-era_2019, mootoovaloo_parameter_2020, Bonici:2022xlo, To:2022ubu, Zhong:2024xuk, Saraivanov:2024soy, Boruah:2024tkq}.

While the acceleration of likelihood evaluations through emulation marks a significant step forward, it is only one facet of the broader efficiency gains that emulators can bring to cosmological analyses. Emulators enable the use of gradient-based samplers, which are particularly well-suited to high-dimensional and complex parameter spaces. These samplers, such as Hamiltonian Monte Carlo (HMC)~\cite{hoffman2011nouturnsampleradaptivelysetting, betancourt2018} and its variants, exploit the differentiability of the model to explore parameter spaces more efficiently than gradient-free methods. As a result, they significantly reduce the number of evaluations required to reach chain convergence, especially for complex target distributions, offering substantial improvements in computational performance.

This capability has led to a growing body of work employing gradient-based samplers in cosmology, with applications to CMB and LSS analyses~\cite{Millea:2020cpw, Li:2022qlf, Campagne:2023ter, piras2023, nygaard_connect_2022, Hahn:2023nvb, Cagliari:2023mkq, capse, Bonici:2022xlo, Ruiz-Zapatero:2023hdf, Mootoovaloo:2024lpv, Giovanetti:2024zce, SPT-3G:2024atg, Zhang:2024thl}. The increasing adoption of these approaches reflects the potential of gradient-based methods to accelerate inference in challenging, high-dimensional models.

The primary method for computing log-likelihood gradients is known as algorithmic differentiation (AD), or automatic differentiation~\cite{griewank2008evaluating, blondel2024, scardapane}. At its core, AD is built on a straightforward concept: any computer program can be interpreted as a sequence of elementary functions, each with a known analytical derivative\footnote{A possible exception is represented by programs with discrete stochastic behavior; in order to deal with this scenario, tailored approaches have been developed as in~\cite{2022arXiv221008572A, Horowitz:2022fvl}.}. The task of an AD engine is to apply the chain rule to combine these derivatives systematically. While AD is capable of differentiating highly complex programs, it can still benefit from higher-level mathematical insights. When custom, hand-written derivatives are provided, significant speed improvements can be achieved. A classic example is matrix multiplication; although this operation can be broken down into individual sums and products, each of which is differentiable, it is far more efficient to express its derivative as another matrix product, which achieves the same result with fewer computational resources~\cite{jurling}.

Several codes have already been developed to efficiently compute quantities relevant for galaxy clustering analyses. Notable examples include \comet{}~\cite{Eggemeier:2022anw}, \matryoshka{}~\cite{Donald-McCann:2022pac}, and \emulatelss{}~\cite{DeRose:2021pqx}. These frameworks have been successfully applied to a variety of cosmological datasets, demonstrating their effectiveness in this field. However, while all these frameworks are differentiable, none of them has been explicitly used in synergy with gradient-based samplers.

In this context, we introduce \effort{}, a \julia{}~\cite{bezanson_julia_2015} package that builds on these previous works with a focus on computational performance and integration with gradient-based algorithms. One of the key strengths of \effort{} is its adaptability: observational effects—such as the Alcock-Paczynski (AP) effect and the survey mask—can be either incorporated into the data generated to train the emulator or included a posteriori. While the former approach is computationally more efficient, as no post-processing of the emulator output is required, the latter is more flexible as the output can be adapted without the need to retrain the emulator. Furthermore, great care has been devoted to ensure \effort{} compatibility with AD engines, enabling efficient differentiation throughout the entire workflow. While previous codes have laid the groundwork, \effort{} offers a distinct advantage for analyses centered on gradient-based techniques, providing a robust and flexible toolkit tailored to the evolving needs of modern cosmological research.

Through its interface with the probabilistic programming language (PPL) \turing{}\footnote{\href{https://turinglang.org/}{https://turinglang.org/}}, \effort{} can be used in conjuction with gradient-based sampling and optimization techniques, making it suitable for computing posteriors in Bayesian analyses  as well as profile likelihoods (and best-fit points) for frequentist constraints.

The design of \effort{} prioritizes computational efficiency without sacrificing accuracy, enabling the rapid exploration of parameter spaces for the analysis of big cosmological datasets. To validate the performance of \effort{}, we applied it to the PT-challenge simulations~\cite{Nishimichi:2020tvu} and the BOSS data~\cite{BOSS:2015npt}, finding no significant deviations from more traditional pipelines that combine \class{} or \camb{} with an EFTofLSS code.

This paper presents the technical details of \effort{}, the methodologies behind its implementation, and its application to both current and future cosmological surveys, with particular emphasis on the upcoming datasets from DESI and Euclid.

This paper is structured as follows: in Sec.~\ref{sec:effort}, we introduce the core features of \effort{}, detailing its architecture, preprocessing strategies, and the implementation of observational effects. Sec.~\ref{sec:inference} describes the probabilistic programming framework and the gradient-based sampling methods employed for cosmological inference. In Sec.~\ref{sec:accuracy}, we validate the performance of \effort{} through applications to the PT-challenge simulations and the BOSS dataset, comparing its results against standard pipelines. Finally, in Sec.~\ref{sec:conclusions}, we present our conclusions and outline potential future developments, including applications to upcoming surveys and extensions of the emulator framework.

\section{The \effort{} Emulator}  
\label{sec:effort}

In this section, we describe the core features of the \effort{} emulator, highlighting its architecture, preprocessing strategies, training approach, and the implementation of observational effects. \effort{} enables fast computation of the galaxy power spectrum at 1-loop in the EFTofLSS (we refer the reader to Appendix~\ref{sec:eftoflss} for a brief review) and seamless integration with AD systems. We discuss its efficient handling of bias parameters, its neural network (NN) architecture, and the methods used to incorporate observational effects, ensuring high precision and computational performance.

\subsection{Neural Network Architecture and Core Functionalities}

\effort{}\footnote{\href{https://github.com/CosmologicalEmulators/Effort.jl}{\texttt{https://github.com/CosmologicalEmulators/Effort.jl}}} is a software package developed in \julia{} \cite{julia}, a high-level, high-performance programming language tailored for technical computing. The choice of \julia{} allows \effort{} to achieve exceptional performance, with the ability to compute the power spectrum multipoles in approximately $15\,\mu\mathrm{s}$\footnote{All the benchmarks of this paper are performed locally using a single core of an i7-13700H CPU.}. Additionally, \julia{}'s support for various AD systems plays a crucial role in enabling efficient sampling, a key feature of the use of \effort{} for cosmological parameter inference.

The key functionalities useful for building emulators, such as commands for the instantiation and execution of trained emulators and postprocessing of the emulators output, are abstracted in an external package, \ace{}\footnote{\href{https://github.com/CosmologicalEmulators/AbstractCosmologicalEmulators.jl}{\texttt{https://github.com/CosmologicalEmulators/AbstractCosmologicalEmulators.jl}}}. This allows us to build other emulators such as \capse{}, a surrogate model for CMB observables~\cite{capse}, with minimal effort.

The \ace{} package integrates two different NN libraries: \scs{}\footnote{\href{https://github.com/PumasAI/SimpleChains.jl}{\texttt{https://github.com/PumasAI/SimpleChains.jl}}} and \lux{}\footnote{\href{https://github.com/LuxDL/Lux.jl}{\texttt{https://github.com/LuxDL/Lux.jl}}}. \scs{} offers superior performance on central processing units (CPUs), while \lux{} extends compatibility to graphics processing units (GPUs).

Moreover, \ace{}’s methods for emulator instantiation via \texttt{JSON} configuration files eliminates the need for pickling or serialization of trained NNs. This approach ensures robustness, portability, and compatibility across different computational environments.

In addition to the NN architecture, \effort{} handles observational effects such as survey window masking and the Alcock-Paczynski (AP) effect. These effects are implemented using tools from the \julia{} ecosystem, ensuring that all steps are compatible with the AD framework.

Thanks to these design choices, \effort{} is easy to train even using standard hardware such as CPUs, making it particularly efficient and accessible, avoiding the need for large-scale hardware setups typically required for more resource-demanding models.

\subsection{Bias Treatment and Emulator Structure}

\effort{} is designed to emulate the galaxy power spectrum multipoles at 1-loop in the EFTofLSS. The computation is carried out as follows:

\begin{equation}
    P_\ell(k; \theta) = \sum_{i,j} b_i b_j \mathcal{P}_{ij, \ell}(k; \theta) + S_{\ell}(k)
\end{equation}
Here, $P_\ell(k; \theta)$ represents the power spectrum multipoles as a function of wavenumber $k$ and cosmological parameters and redshift $\theta$, which control the cosmological model. The terms $\mathcal{P}_{ij,\ell}(k; \theta)$ are the individual components of the power spectrum multipoles, $b_i$ are biases and counterterms, $S_{\ell}(k)$ represents the stochastic terms. By treating biases and counterterms analytically, \effort{} (as \pybird{} does) decouples the elements depending on cosmological parameters from the other ones, significantly reducing the dimensionality of the emulated space. Regarding the stochastic part, this is constructed analytically by the code and does not require to be emulated.

Other packages, such as \emulatelss{}~\cite{DeRose:2021pqx}, incorporate bias parameters directly into the emulation process, which can lead to the need for more complex NNs. In contrast, \effort{} follows a strategy similar to \comet{}~\cite{Eggemeier:2022anw} and \matryoshka{}~\cite{Donald-McCann:2022pac}, where bias parameters are handled analytically. This reduces the complexity of the NNs, allowing for faster training and inference, while the small computational overhead of bias inclusion remains negligible.

\subsection{Preprocessing Strategy}
\label{sec:preprocessing}

\effort{} employs a carefully optimized preprocessing pipeline to ensure both accuracy and efficiency in the emulator. The linear matter power spectrum is given by:

\begin{equation}
    P_\mathrm{lin}(k,z) = A_\mathrm{s} k^{-3} \left(\frac{k}{k_0}\right)^{n_\mathrm{s}-1}T^2(k,z)D^2(z),
\end{equation}
where $A_\mathrm{s}$ and $n_\mathrm{s}$ set the initial power spectrum conditions, $k_0$ is the pivot wavenumber, $T(k,z)$ is the matter transfer function, and $D(z)$ is the scale-independent growth factor.

The $P_{11}$ and counter-term, $P_{ct}$, components are linear in $P_\mathrm{lin}(k,z)$, and therefore scale proportionally with $\mathcal{A}(z)=A_\mathrm{s}D^2(z)$, while the loop component is quadratic in $P_\mathrm{lin}(k,z)$, making it proportional to $\mathcal{A}^2(z)$, up to the infrared (IR) resummation~\cite{Senatore:2014via, Baldauf:2015xfa} and the effects of massive neutrinos. To account for this structure, we rescale the $P_{11}$ and $P_{ct}$ terms by $\mathcal{A}(z)$, and the loop term by $\mathcal{A}(z)^2$.

While this rescaling significantly reduces the complexity, it is not perfect due to the IR resummation. To address these residuals, \effort{} combines the analytical rescaling with machine learning, allowing the NNs to learn the remaining non-factorizable dependencies on $A_\mathrm{s}D(z)$. This hybrid approach ensures high precision with negligible computational overhead. As shown in~\cite{capse}, this method enhances the emulator’s accuracy.

This rescaling is particularly effective for extended models with additional parameters, which primarily influence the amplitude of the linear matter power spectrum (e.g. modifications to the dark energy equation of state), which primarily influence the amplitude of the linear matter power spectrum. The analytical rescaling efficiently captures their impact on the background evolution.

In \comet{} a similar method is used, known as \textit{evolution mapping}, which exploits parameter degeneracies that primarily affect the amplitude of the linear power spectrum~\cite{Sanchez:2021plj}. While \comet{}'s approach is used to reduce the number of input features, \effort{} leverages it primarily to reduce the dynamic range of the output features. This is especially advantageous in cases with scale-dependent growths, such as when accounting for massive neutrinos, where \effort{}'s NN learns the residual effects not captured by the rescaling, offering flexibility in modeling non-linear phenomena. The impact of this approach is assessed and shown in Sec.~\ref{sec:accuracy}.

The growth factor $D(z)$ is thus used to postprocess the output of the NNs, and its computation will be described in the next subsection.

Finally, to enhance numerical stability, \effort{} applies min-max normalization to both input and output features. This ensures consistent scaling across all features, preventing any feature from dominating due to differences in magnitude, which leads to a more stable and efficient NN training process, resulting in improved overall performance~\cite{ioffe2015}.

\subsection{Observational Effects}
\label{sec:observational_effects}

In this subsection, we describe the treatment of key observational effects in \effort{}, including the computation of required quantities such as the growth factor, the inclusion of the Alcock-Paczynski (AP) effect, and the window mask convolution.

\subsubsection{Computation of additional Quantities}
\label{sec:background_quantities}

The accurate calculation of background quantities is fundamental for interpreting large-scale structure data. In \effort{}, we compute several key cosmological background quantities: the Hubble factor $H(z)$, the radial comoving distance $r(z)$, the growth factor $D(z)$, and the growth rate $f(z)$. These quantities are essential inputs for the inclusion of the AP effect (and, in the case of $f(z)$, redshift-space distortions), as we will discuss in the next subsection.

There are two main approaches to handling background quantities in cosmological analyses. One approach, adopted by packages such as \matryoshka{} and \cosmopower{}~\cite{SpurioMancini:2021ppk}, is to emulate the background quantities. While this can speed up calculations, it requires an additional emulator and a well-trained model to ensure precision across the parameter space.

In contrast, \effort{} computes these background quantities dynamically by integrating the relevant equations at runtime. This approach avoids the need for additional emulators, simplifying the pipeline and reducing dependency on pre-computed models. With careful implementation, these calculations can be performed efficiently, providing precise results in a short amount of time.

When computing the Hubble factor, \effort{} accounts for contributions from non-relativistic matter, evolving DE, radiation, and massive neutrinos. In particular, for massive neutrinos, we incorporate the transition from relativistic to non-relativistic regimes, following the treatment described in~\cite{Zennaro:2016nqo, Bayer:2020tko}; here we review the most important equations for completeness.
The dimensionless Hubble factor $E(z) = H(z)/H_0$ is given by

\begin{equation}
E^2(z)=\Omega_{\gamma, 0} (1+z)^{4}+\Omega_{c, 0} (1+z)^{3}+\Omega_\nu(z)+\Omega_{\mathrm{DE}, 0}(1+z)^{3\left(1+w_0+w_a\right)} \exp{\frac{-3 w_a z}{1+z}}.
\end{equation}
\(\Omega_{\gamma, 0}\), \(\Omega_{c, 0}\), and \(\Omega_{\mathrm{DE}, 0}\) are the abundances of radiation, non-relativistic matter, and DE, respectively, and the DE equation of state is parametrized as $w(a) = w_0 + w_a (1-a)$~\cite{Chevallier:2000qy, Linder:2002et}.

The contribution from massive neutrinos is given by

\begin{equation}
\Omega_\nu(a) = \frac{15}{\pi^4} \Gamma_\nu^4 \frac{\Omega_{\gamma, 0}}{a^4} \sum_{j=1}^{N_\nu} \mathcal{F}\left(\frac{m_j a}{k_B T_{\nu, 0}}\right),
\end{equation}
where \(m_j\) is the mass of each neutrino eigenstate, and \(T_{\nu, 0}\) is the current temperature of the neutrino background,  $\Gamma_\nu \equiv T_{\nu, 0} / T_{\gamma, 0}$ is the neutrino-to-photon temperature ratio today and the function \( \mathcal{F}(y) \) is defined as

\begin{equation}
\mathcal{F}(y) \equiv \int_0^{\infty} d x \frac{x^2 \sqrt{x^2+y^2}}{1+e^x}.
\end{equation}
The parameter $\Gamma_\nu$ takes into account corrections to the instantaneous decoupling and it is given by
\begin{equation}
\Gamma_\nu^4 = \frac{1}{3} N_{\text {eff }} \left(\frac{4}{11}\right)^{4 / 3} \, ,
\end{equation}
where we take \(N_{\text{eff}} = 3.044\)~\cite{Froustey:2020mcq, Bennett:2020zkv}.

The comoving distance $r(z)$ is computed as
\begin{equation}
r(z) = c \int_0^z \frac{dz'}{H(z')}.
\end{equation}

We implement two numerical approaches to solve the integral for $r(z)$. First, we employ an adaptive integration Gaussian quadrature scheme, which ensures a high-precision calculation but is more computationally expensive. This method is ideal for cases requiring extremely accurate results. Second, we use a Gauss-Lobatto rule with a fixed number of points (9), which provides sufficient precision for most practical purposes, yielding a relative accuracy better than $10^{-5}$ compared to the adapative quadrature scheme\footnote{The adaptive Gaussian quadrature scheme is used within our suite of unit tests to verify that the faster Gauss-Lobatto method is accurate enough for typical cosmological applications.}. We further validated our results with \camb{} and \texttt{pyccl}~\cite{LSSTDarkEnergyScience:2018yem}, finding an agreement up to the fifth digit.

The growth factor $D(a)$ is computed by solving the following second-order differential equation:
\begin{equation}
D^{\prime \prime}(a) + \left(2 + \frac{E^{\prime}(a)}{E(a)}\right) D^{\prime}(a) = \frac{3}{2} \Omega_\mathrm{m}(a) D(a) \, ,
\end{equation}
where $D^{\prime} \equiv \frac{d D}{d \ln a}$.

We start solving this equation at high redshift, nominally $z = 138$, deep in the matter domination regime. The appropriate initial conditions in this regime are given by:
\begin{equation}
\begin{aligned} 
D\left(z_i\right) &= a_i \\
D^{\prime}\left(z_i\right) &= a_i,
\end{aligned}
\end{equation}
where $a_i$ corresponds to the initial scale factor.

To integrate this equation, we use the Tsitouras algorithm~\cite{TSITOURAS2011770}, a Runge-Kutta method known for its efficiency and accuracy.

The solver is implemented in the \odejl{}\footnote{\href{https://github.com/SciML/DifferentialEquations.jl}{https://github.com/SciML/DifferentialEquations.jl}} package, part of the \julia{} ScientificMachineLearning (SciML) suite, which provides a high-performance environment for solving differential equations~\cite{rackauckas2017differentialequations}.

The solver is made fully differentiable, supporting both forward and backward automatic differentiation (AD) systems. This ensures that gradients can be computed accurately and efficiently during optimization processes that require differentiation of the solver output, such as when integrating \effort{} with gradient-based pipelines.

\begin{figure}[htbp]
\centering
\includegraphics[width=
\textwidth]{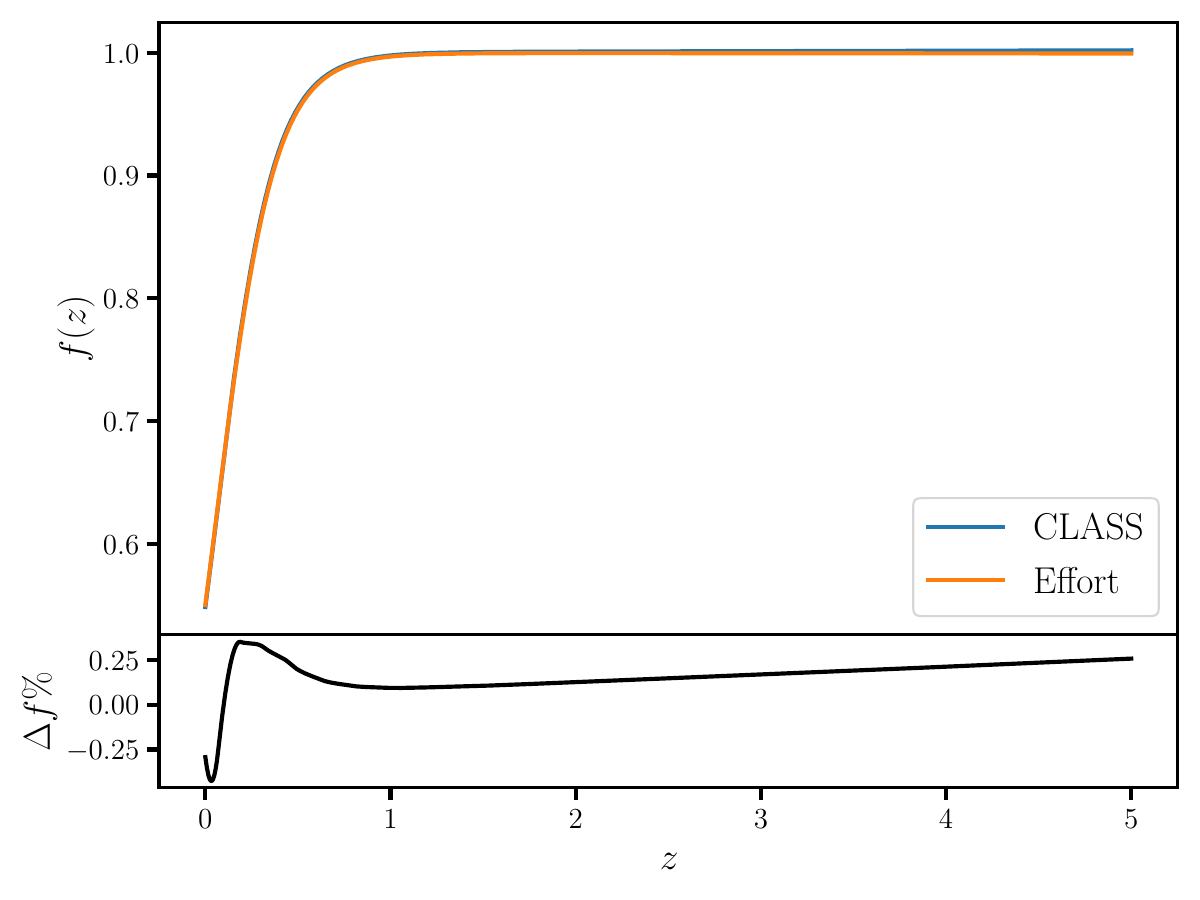}
\caption{Comparison of the growth rate $f(z)$ as  computed by \class{} and \effort{}. The two codes have a remarkably good agreement, with differences smaller than $1\%$ over the entire redshift range considered. \label{fig:growth} }
\end{figure}

The growth rate $f(z)$ is related to the evolution of the growth factor through
\begin{equation}
f(z) \equiv \frac{d\ln D(z)}{d\ln a},
\end{equation}
where $a$ is the scale factor. The solver also retrieves the first derivative of the growth factor, \( D^{\prime}(a) \), which can be used directly to compute $f(z)$. These background quantities are dynamically computed for each cosmological model considered in \effort{}, providing a flexible and precise foundation for modeling galaxy clustering. A comparison between the growth rate computed by \class{} and \effort{} is shown in Fig.~\ref{fig:growth}; for this test, we considered a cosmological model with a single massive neutrino with $M_\nu=0.5\,\mathrm{eV}$ and an extreme scenario for the DE equation of state, with the values for $w_0$ and $w_a$ being respectively of $-2$ and $-3$.

\subsubsection{Alcock-Paczynski (AP) Effect}
\label{sec:ap_effect}

The Alcock-Paczynski (AP) effect is a geometric distortion that arises when converting redshift space into physical space, since one must assume a reference cosmological model~\cite{alcock}. This effect causes anisotropies in the observed galaxy clustering, affecting the inferred distances along and across the line-of-sight. In this work, we follow the notation used in~\cite{DAmico:2020kxu, Ivanov:2019pdj}.

To account for the AP effect in \effort{}, we calculate the distortion factors as follows~\footnote{These are the appropriate definitions when using wavenumber units of $h/\textrm{Mpc}$, which we assume throughout.}:

\begin{equation}
    q_{\|}=\frac{D_A(z) H(z=0)}{D_A^{\text {ref }}(z) H^{\text {ref }}(z=0)}, \quad q_{\perp}=\frac{H^{\mathrm{ref}}(z) / H^{\mathrm{ref}}(z=0)}{H(z) / H(z=0)}
\end{equation}

Here, \(q_{\|}\) and \(q_{\perp}\) represent the distortions along and transverse to the line-of-sight, respectively. These scaling factors adjust for the differences between the reference and true cosmology, where \(D_A(z)\) is the angular diameter distance, and \(H(z)\) is the Hubble factor.

The transformation of the power spectrum multipoles due to the AP effect is given by:

\begin{equation}
    P_{\ell}(k^{\text {ref}})=\frac{2 \ell+1}{2 q_{\|} q_{\perp}^2} \int_{-1}^1 \mathrm{~d} \mu^{\text {ref }} P\left(k\left(k^{\text {ref }}, \mu^{\text {ref }}\right), \mu\left(\mu^{\text {ref }}\right)\right) \mathcal{L}_{\ell}\left(\mu^{\text {ref }}\right).
\end{equation}

This equation expresses the power spectrum multipoles \(P_{\ell}(k)\), where \(\mu^{\text {ref }}\) is the cosine of the angle between the wavevector \(k\) and the line-of-sight in the reference cosmology. The function \(\mathcal{L}_{\ell}(\mu^{\text {ref }})\) is the Legendre polynomial of order \(\ell\), and the transformation adjusts the power spectrum to account for differences between the true and reference cosmology.

The relations between the wavevector \(k\) and angle \(\mu\) in the true and reference cosmologies are given by:

\begin{equation}
    k=\frac{k^{\text {ref }}}{q_{\perp}}\left[1+\left(\mu^{\text {ref }}\right)^2\left(\frac{1}{F^2}-1\right)\right]^{1 / 2}, \quad \mu=\frac{\mu^{\text {ref }}}{F}\left[1+\left(\mu^{\mathrm{ref}}\right)^2\left(\frac{1}{F^2}-1\right)\right]^{-1 / 2}
\end{equation}

Here, \(F=q_{\|} / q_{\perp}\) is the anisotropy factor, capturing the relative stretching between the line-of-sight and transverse directions due to cosmological distortions. These transformations ensure that the clustering signal is correctly interpreted in terms of the true cosmological model.

We implemented two methods to integrate the AP effect. We use an adaptive Gaussian quadrature scheme for high precision, although this approach is more computationally expensive. Alternatively, we use a Gauss-Lobatto quadrature with 5 points, which achieves near floating-point precision with reduced computational cost.

The main challenge arises when incorporating AD. To compute the AP effect, we need to interpolate the emulator’s output onto a new \(k\)-grid, whose specific values depend on cosmological parameters. For this, we employ a quadratic spline interpolation, as in \pybird{}. However, calculating the Jacobians of the splines with respect to their input leads to a sparse Jacobian matrix, which the AD system does not naturally optimize for.

Although the AP calculation itself takes approximately $30\,\mu\mathrm{s}$, backward propagation of gradients through the AP calculation initially took around $100\,\mathrm{ms}$, significantly degrading \effort{}'s performance. 

To resolve this, we reimplemented the spline interpolation and wrote custom backward differentiation rules that take advantage of the sparsity pattern. This optimization reduced the time for calculating the Jacobians to approximately $200\,\mu\mathrm{s}$, representing a three-order-of-magnitude improvement.

We validated the correctness of these custom differentiation rules using both forward-mode AD and finite difference methods.

\subsubsection{Window Mask Convolution}
\label{sec:window_mask}

The finite size and irregular geometry of surveys introduce an observational window function that modulates the observed clustering signal. Accounting for this requires a multiplication of the theory correlation function with a window mask in real space, or a convolution of the window mask with the theory power spectrum in Fourier space.

In \effort{}, we adopt the approach of~\cite{Beutler:2018vpe}, implementing the window convolution as an efficient array contraction. The window function, derived from the survey mask, is applied to the theoretical predictions at each step of the MCMC sampling, ensuring that observational effects are consistently incorporated into the analysis.

The convolution is executed using \tullio{}~\cite{michael_abbott_2023_10035615}, a high-performance framework for array operations. The entire process is completed in a few $\mu\mathrm{s}$. To further enhance performance, we have written custom differentiation rules, enabling the automatic differentiation (AD) system to efficiently manage the convolution operation.

By accurately incorporating these observational effects, \effort{} delivers precise predictions for the power spectrum, supporting robust cosmological parameter inference.

\section{Cosmological Inference Framework} 
\label{sec:inference}

\subsection{Probabilistic Programming with \turing{}}

For cosmological parameter inference, we employ a probabilistic programming language (PPL) which is built in \julia{}, namely \turing{}~\cite{ge2018t}. PPLs allow users to specify probabilistic models that include the necessary priors and likelihoods, streamlining the process of drawing samples from the posterior distribution. This flexibility is particularly useful for complex models, such as those involving cosmological parameters, as it enables integration with advanced samplers and AD frameworks.

\turing{} provides a straightforward way to define probabilistic models by allowing users to declare their priors and construct the likelihood function. Once the model is defined, \turing{} takes care of deriving the posterior and can utilize various sampling algorithms for inference. A key strength of \turing{} is its compatibility with gradient-based samplers, such as the Hamiltonian Monte Carlo (HMC) and its variants, including the No-U-Turn Sampler (NUTS) and MicroCanonical Langevin Monte Carlo~\cite{Robnik:2023pgt}.

In this work, we use NUTS and MicroCanonical Langevin Monte Carlo~\cite{Robnik:2023pgt} as samplers, all of them accessed through the \turing{} interface. This approach has been successfully employed in various cosmological studies, as demonstrated in~\cite{capse, Ruiz-Zapatero:2023hdf, Bonici:2022xlo}. These samplers efficiently explore the posterior distributions of cosmological and nuisance parameters, taking full advantage of the differentiability of \effort{}.

\subsection{The Hamiltonian MonteCarlo sampler}
\label{sec:hmc}
The Hamiltonian Monte Carlo (HMC), also known as Hybrid Monte Carlo, is a state-of-the-art method for drawing samples from a probability distribution~\cite{betancourt2018}. It is particularly useful when the distribution is over high-dimensional spaces and/or with complex geometries.

The key idea behind HMC is to introduce momentum variables, $p$, for each parameter in the model, which is instead represented as a position variable $q$. A potential energy $V(q)$ is introduced, given as minus the logarithm of the joint-likelihood $\log L$:
\begin{equation}
    -\log \mathcal{P}(q)=V(q) \quad H(q, p)=V(q)+U(q, p),
\end{equation}

where the kinetic term is given by:
\begin{equation}
    U(q, p)=p^T M^{-1} p.
    \label{eq:kin}
\end{equation}

These momentum variables, together with the original parameters, form a Hamiltonian system. The system evolves according to Hamilton’s equations, which in the vanilla case read:
\begin{equation}
    \begin{aligned}
& \frac{\mathrm{d} p}{\mathrm{~d} t}=-\frac{\partial V}{\mathrm{~d} q}=\frac{\partial \log \mathcal{P}}{\mathrm{d} q} \\
& \frac{\mathrm{d} q}{\mathrm{~d} t}=+\frac{\partial U}{\mathrm{~d} p}=M^{-1} p.
\end{aligned}
\end{equation}

In the vanilla HMC, the system evolves deterministically for a fixed amount of time, following a trajectory that is likely to stay in regions of high probability under the target distribution. The trajectory is usually numerically integrated with a symplectic integrator.

The deterministic evolution in HMC allows it to explore the target distribution more efficiently. However, it requires the ability to compute gradients of the log-probability of the target distribution, which can be computationally expensive for complex models.

Vanilla HMC is a powerful sampling method that combines ideas from physics and statistics and can efficiently draw samples from high-dimensional distributions, but requires tuning and may be computationally expensive for complex models. One of the main challenges with the vanilla HMC is the need to carefully choose and tune the step size and the number of steps. If these parameters are not set correctly, the HMC can either miss important regions of the parameter space or waste computational resources by taking too many steps.

The No-U-Turn Sampler (NUTS) is a self-tuning variant of the Hamiltonian Monte Carlo (HMC) method, designed to address some of the shortcomings of the vanilla HMC~\cite{hoffman2011nouturnsampleradaptivelysetting}. The name “No-U-Turn” comes from the sampler’s unique feature of stopping its trajectory once it starts to turn back on itself, hence avoiding unnecessary computations.

NUTS tunes HMC hyper-parameters by adaptively choosing the trajectory length. It starts with a trajectory of length 1, and then doubles the length as long as the trajectory does not make a U-turn. A U-turn is detected when the current position and the proposed new position are moving in opposite directions along the gradient of the log-probability.

This adaptive algorithm ensures that NUTS spends more time exploring new regions of the parameter space and less time retracing its steps. It also removes the need for manual tuning of the trajectory length, making NUTS more user-friendly than vanilla HMC.

However, like HMC, NUTS requires the ability to compute gradients of the log-probability, which can be computationally expensive for complex models. In our case, the computation of the log-likelihood gradient is performed using one of the \julia{} AD systems.

\subsection{MicroCanonical Hamiltonian Monte Carlo Sampler}
\label{sec:mchmc}
The MicroCanonical Hamiltonian Monte Carlo (MCHMC) sampler~\cite{Robnik:2022bzs, Robnik:2023pgt}, is a variation of the traditional HMC method. Unlike standard HMC, which relies on a Metropolis-Hastings correction step to ensure detailed balance, MCHMC operates within the microcanonical ensemble, where the Hamiltonian energy $H(q, p)$ is conserved throughout the entire trajectory. Removing the need for Metropolis corrections, reduces computational overhead and increases sampling efficiency.

MCHMC also employs a variable mass Hamiltonian in Eq.~\eqref{eq:kin}, which adapts to the geometry of the posterior distribution. In regions of higher probability density, the particle moves slower, improving the sampler’s capacity to explore complex and high-curvature posterior distributions.

This method is especially well-suited for cosmological inference tasks, where the parameter space is often large and complex. MCHMC has been demonstrated to outperform traditional HMC methods in terms of sampling efficiency for such high-dimensional problems, providing faster convergence while maintaining accuracy in the estimation of posterior distributions~\cite{Robnik:2023pgt, capse, Bayer:2023rmj}.

\section{Results}
\label{sec:results}
\subsection{Accuracy checks and preprocessing impact}
\label{sec:accuracy}
Having laid out the structure of \effort{}, we can now discuss a concrete implementation.
Given its flexible structure, it is possible to use \effort{} with two different approaches, \textit{i.e.} including redshift as a free parameter or working at a specific redshift. While the former approach is more flexible, as it enables us to train one generic emulator that can be used in more situations, the latter option naturally leads to a more precise surrogate, as we are removing one of the input parameters. Considering that surveys like DESI and Euclid divide galaxy in only a handful of bins with a specific effective redshift, the second approach is useful in case one needs very high precision. Here we focus on an emulator of the former kind, since we want to provide a tool that does not require additional training. The training dataset contained 60,000 couples of input parameters and $P(k)$s. We input to the NN the redshift and cosmological parameters distributed according to the Latin Hypercube in the parameter range described in Table~\ref{tab:parameter_bounds}. We employ NNs with 5 hidden layers, each of them with 64 neurons. We used a batch size of 128, and trained for 100,000 epochs using an ADAM optimizer with an initial learning rate of $10^{-4}$; when using an 8-core CPU, this process took around one hour of wall-time. We employed a vanilla mean square error loss function for the emulator presented in this section and used in the BOSS analysis, while for the PT-challenge we used a modified loss function which included a $1/k^2$ weight. The dataset as been split in $80\%$ and $20\%$ for, respectively, the training and test sets. To prevent overfitting, we save the weights only when the loss on the testing dataset shows improvements.

\begin{table}[h!]
    \centering
    \begin{tabular}{|c|cc|}
        \hline
        Parameter & Min & Max \\ \hline
        $z$ & 0.25 & 2.0 \\ \hline
        $\ln 10^{10}A_\mathrm{s}$ & 2.5 & 3.5 \\ \hline
        $n_\mathrm{s}$ & 0.8 & 1.10 \\ \hline
        $H_0$ [km/s/Mpc] & 50.0 & 80.0 \\ \hline
        $\omega_{b}$ & 0.02 & 0.025 \\ \hline
        $\omega_{c}$ & 0.08 & 0.2 \\ \hline
        $M_\nu$ [eV] & 0.0 & 0.5 \\ \hline
    \end{tabular}
    \caption{Parameter bounds for redshift and cosmological parameters.}
    \label{tab:parameter_bounds}
\end{table}

Before showing a few applications of trained emulators, we want to gauge the impact of the preprocessing procedure outlined in Sec.~\ref{sec:preprocessing}.

\begin{figure}[ht!]
  \centering
  \includegraphics[width=1\textwidth]{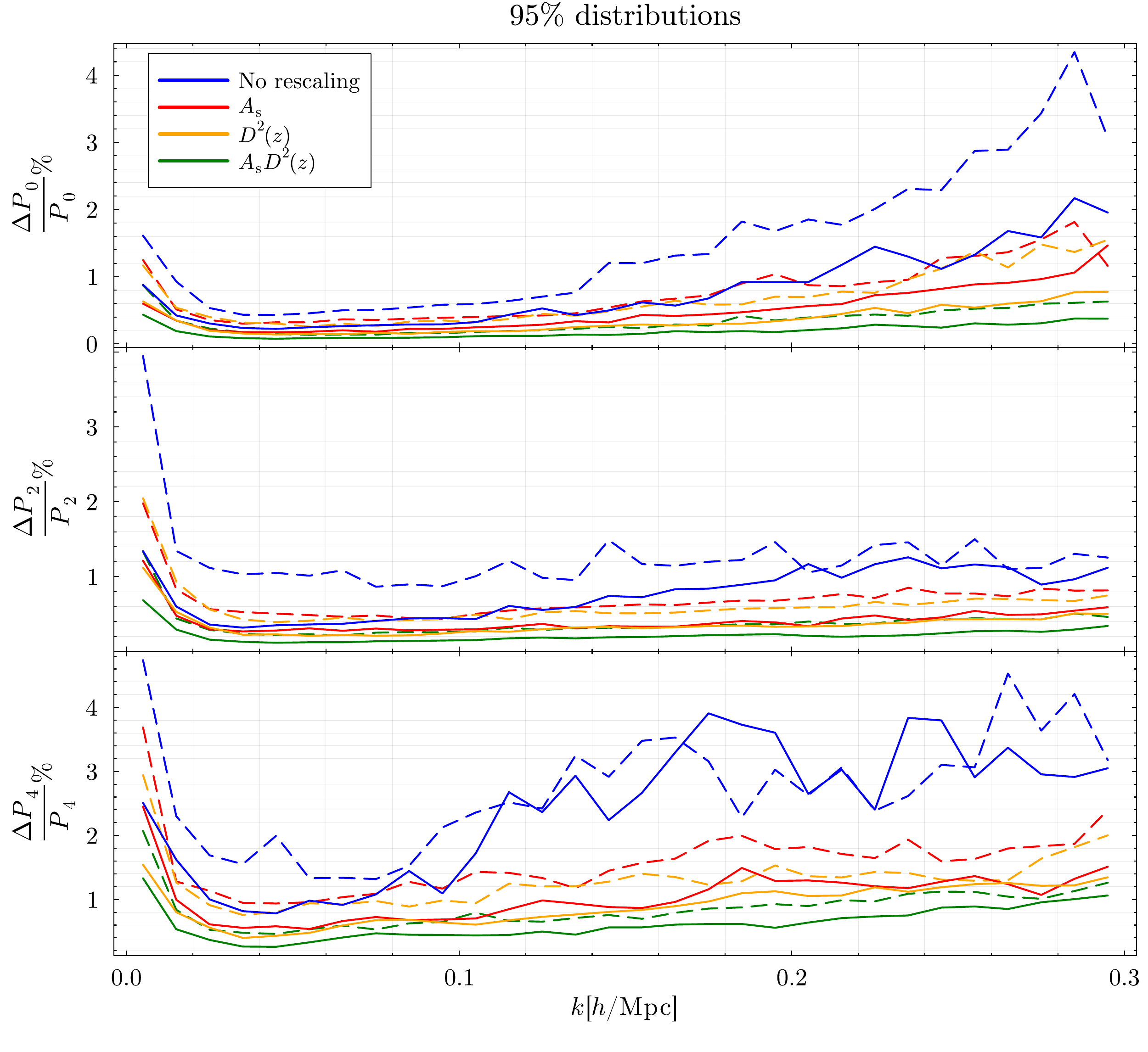} 
  \caption{95\% distribution of the percentage errors. We show the impact of the different rescaling schemes employed. The dot-dashed and solid lines refers respectively to a training dataset with 20,000 and 60,000 elements. As can be seen, a bigger dataset helps in enhancing the performance, but the rescaling is even more impactful. For this test, we fix the EFT parameters, using a combination obtained from the measurements employed in~\cite{Zhang:2024thl}.}
  \label{fig:preprocessing_residuals}
\end{figure}

The question we aim to answer is: how can we improve the performance of our emulators? The straightforward approach is to use a bigger dataset and/or a bigger neural network, according to the approximation theorem~\cite{Cybenko1989ApproximationBS}. The second approach which we want to investigate here is a rescaling of the output features which depends on the value of the input features; this method has already been applied in~\cite{capse}, where the CMB spectra where rescaled by $A_s$, leading to a reduction of the residuals by a factor of 3, clearly showing the benefit of this strategy. Here we can make a further step, dividing the output features by the growth factor as outlined in Sec.~\ref{sec:preprocessing}.

In order to assess the impact of rescaling on the observables considered here in a quantitative way, we study the distribution of the percentage residuals in the following 4 scenarios:
\begin{itemize}
    \item No rescaling is employed
    \item We rescale by $A_\mathrm{s}$
    \item We rescale by $D^2(z)$
    \item We rescale by $A_\mathrm{s} D^2(z)$
\end{itemize}
Furthermore, we consider two training datasets, with 20,000 and 60,000 samples. The results of this comparison are shown in Fig.~\ref{fig:preprocessing_residuals}. As can be seen from the comparison of the dashed and solid lines, which represents respectively the 20,000 and the 60,000 samples training dataset, getting a bigger dataset helps in improving the accuracy of the emulator. However, we want to emphasize that the rescalings have a much bigger impact. This shows how it is possible to get a high-precision emulator even when using a small neural-network: in other studies, like~\cite{SpurioMancini:2021ppk, Donald-McCann:2022pac, Trusov:2024mmw}, the NNs employed had a significantly higher number of neurons.

Finally, it is worth mentioning that our most effective physics-based preprocessing procedure, while lowering the emulation error, relies on the solution of an ODE that, although relatively fast (on the order of $150\,\mu\mathrm{s}$), has now become the main bottleneck of our pipeline since 
\effort{} computes each multipole in around $15\,\mu\mathrm{s}$. To circumvent this issue, we employed a symbolic regression approach (using \texttt{SymbolicRegression.jl}~\cite{symbolic_regression}) to emulate the ODE solution by directly seeking a closed-form expression for the growth factor. This strategy offers two main advantages: first, once the symbolic expression is found, it can be ported to any programming language without loss of precision; second, it greatly speeds up the preprocessing step. Indeed, evaluating the symbolic expression takes only $200\,\mathrm{ns}$, offering a dramatic speedup compared to the ODE solution (see also~\cite{Bartlett:2023cyr, Bartlett:2024jes, Sui:2024wob} for other applications of symbolic regression related to LSS observables).

\begin{figure}[ht!]
  \centering
  \includegraphics[width=1\textwidth]{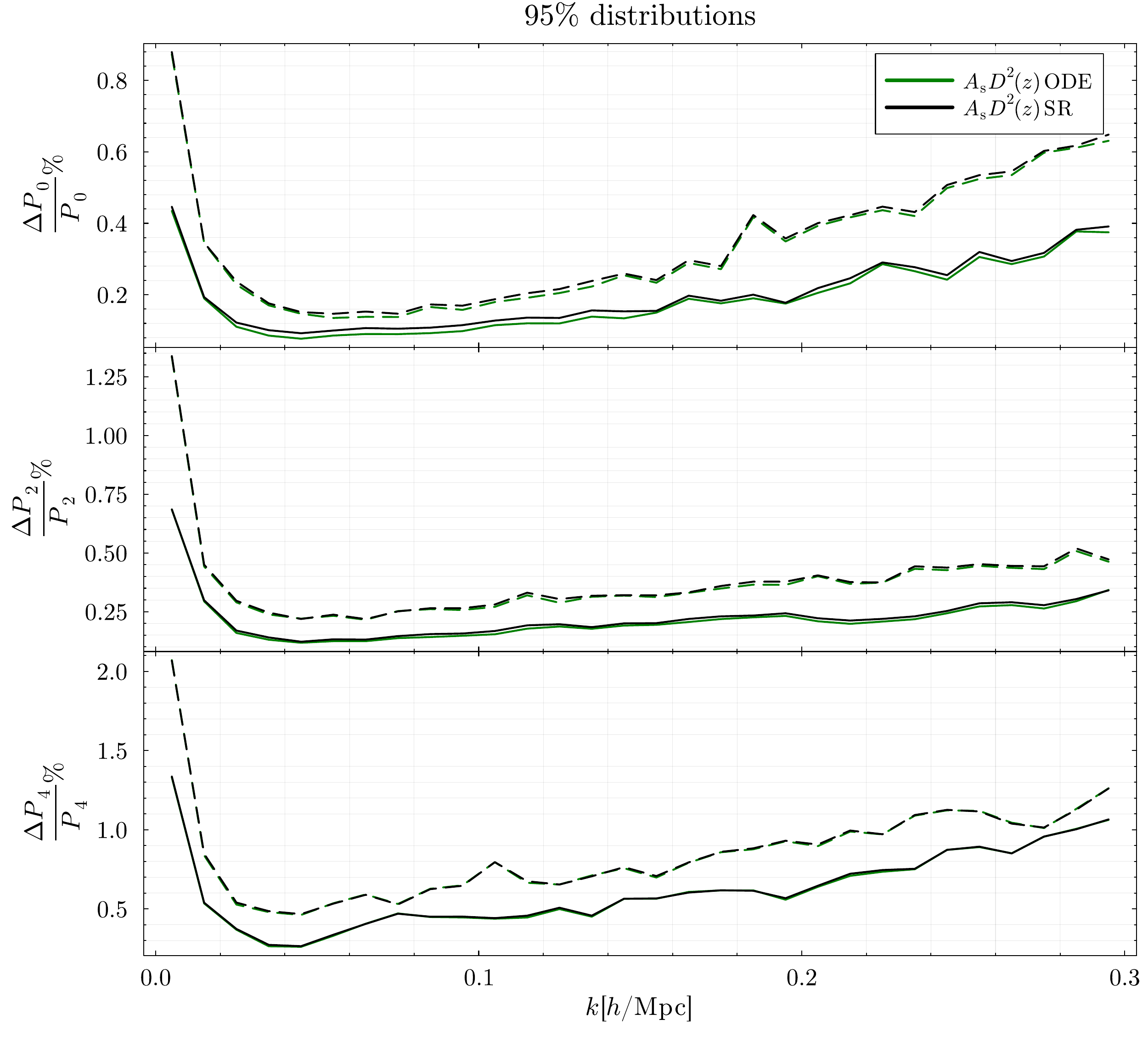} 
  \caption{95\% distribution of the percentage errors, comparing the ODE and the symbolic regression rescaling. As can be seen, the two approaches give almost indistinguishable results, with the sole exception of the monopole emulator trained with 60,000 samples, as in that case the growth factor symbolic emulator error is not negligible compared to the ODE rescaled emulator error.}
  \label{fig:preprocessing_residuals_sr}
\end{figure}

Moreover, while joint analyses can partially amortize the ODE cost by solving it once for multiple redshifts, it remains desirable to reduce this expense further. In our tests, the symbolic expression we found is accurate at the 0.1\% level in the same parameter range explored by our trained 
\effort{} emulator for 99.87\% of the validation dataset. Substituting the ODE solver with the symbolic growth factor does not affect the final emulator performance, except in the specific case of the monopole emulator trained on 60,000 samples, where we observe a modest increase in the residual error, which is still comfortably below 0.4\%, as can be seen from Fig.~\ref{fig:preprocessing_residuals_sr}. Nonetheless, the symbolic approach proves advantageous in terms of portability, consistency, and computational efficiency. However, for users who do not want to sacrifice accuracy, it is still possible to choose to use the ODE solver in \effort{}.

Supported by our findings, we advocate for physics-based preprocessing, as it helps in improving the accuracy of the emulators by leveraging the domain knowledge of the scientists.

\subsection{Application to the PT-challenge simulations and BOSS}
\label{sec:data_applications}
In this section we show the application of \effort{} to two different sets of measurements, the PT-challenge and BOSS datasets.

The PT-challenge\footnote{\href{https://www2.yukawa.kyoto-u.ac.jp/~takahiro.nishimichi/data/PTchallenge/}{https://www2.yukawa.kyoto-u.ac.jp/~takahiro.nishimichi/data/PTchallenge/}} was designed to demonstrate the reliability and robustness of the EFTofLSS as a powerful and trustworthy tool for analyzing cosmological datasets. To achieve this, a set of high-precision simulations was generated. This approach provided an ideal testing ground for validating the accuracy and effectiveness of theoretical models like the EFTofLSS.

The simulations consisted of ten boxes with independent random realizations, each comprising \(3{,}072^3\) mass elements within a periodic cube of side length \(3{,}840\, h^{-1}\)~Mpc, resulting in a total simulated volume of \(566\, (h^{-1}\, \text{Gpc})^3\). Initial conditions were generated with second-order Lagrangian Perturbation Theory (2LPT) and evolved using \textsc{Gadget2}. Particle snapshots were recorded at six redshifts: $z =$ 3, 2, 1, 0.61, 0.51, and 0.38.

Halos were identified with the \textsc{Rockstar} halo finder, and galaxies were probabilistically assigned based on the virial mass of halos, ensuring that mock galaxies were consistent with observational data.

The resulting mock galaxy catalogs correspond to redshifts \(z = 0.38\) (LOWZ), \(z = 0.51\) (CMASS1), and \(z = 0.61\) (CMASS2), mirroring the observational data sets; specifically, the measurement employed in this work are those at $z=0.61$. These simulations provide a robust framework for testing cosmological models and statistical methods, owing to their large volume and the precision of the mock galaxy distributions\footnote{The \julia{} version of the PT-challenge likelihood is available \href{https://github.com/JuliaCosmologicalLikelihoods/BlindedChallenge.jl}{here}.}. Here we analyzed these simulations up to $k_\mathrm{max}=0.12\,h/\mathrm{Mpc}$.

The Baryon Oscillation Spectroscopic Survey (BOSS) provides publicly available redshift catalogs that enable galaxy clustering measurements~\cite{2013AJ....145...10D, BOSS:2016wmc}.
Power spectrum multipoles and window function matrices are the ones used in~\cite{DAmico:2022osl}.

The BOSS dataset is divided into subsamples based on the Northern and Southern Galactic Caps (NGC and SGC). We divide them in two redshift bins (CMASS and LOWZ), corresponding to effective redshifts $z_{\text{eff}} = [0.57, 0.32]$ respectively. This results in a total of four sets of multipoles\footnote{The \julia{} version of the BOSS likelihood is available \href{https://github.com/JuliaCosmologicalLikelihoods/BOSSLikelihoods.jl}{here}.}.
We fit the power spectra with $k_\mathrm{max} = 0.23\,h/\mathrm{Mpc}$ for CMASS and $0.20\, h/\mathrm{Mpc}$ for LOWZ.

We validate the accuracy of our emulators on these datasets by comparing the Bayesian posteriors produced by \effort{} and \pybird{}. Specifically, we present the posterior distributions for the cosmological and EFT parameters that \pybird{} cannot marginalize analytically, although \effort{} does not perform the analytical marginalization. To preserve the secrecy of the true cosmology of the PT-challenge, axis ticks and labels are omitted from the plots related to that analysis. 

Regarding the prior, we used the same prior and analysis settings used in~\cite{Nishimichi:2020tvu} from the West Coast Team and we developed a different emulator, tailored to have only 3 input cosmological parameters~\footnote{We will not release the trained emulators for this part of the work to ensure the input cosmology remains undisclosed to the community.} The results demonstrate remarkable accuracy, with an excellent agreement between the two methods, consistent with Montecarlo noise. In terms of performance, the \pybird{} chains were generated using the \montepython{} sampler~\cite{Brinckmann:2018cvx} and required several hours to obtain a very high degree of convergence on multiple CPUs of a computing cluster. By contrast, \effort{} achieved convergence with significantly greater efficiency. We ran four chains for both the NUTS and MCHMC samplers, using 500 and 10,000 burn-in steps and 2,000 and 200,000 accepted steps, respectively. These chains were initialized using \texttt{Pathfinder.jl}, a variational inference method that quickly generates samples from the typical set~\cite{2022arXiv221008572A, seth_axen_2024_14576747}. The entire analysis with \effort{} required approximately 10 minutes on a laptop and converged to the same posterior distribution as \pybird{}. The Effective Sample Size per second (ESS/s) was 1.2 for NUTS and 5.1 for MCHMC.
We stress that, given the large volume of the PT-challenge, this is a very stringent test of the accuracy our emulator.

A direct comparison of sampling efficiency with \pybird{} was not made, as \pybird{} samples from a smaller-dimensional parameter space. However, even when analyzing a higher-dimensional space, \effort{} achieves a sampling efficiency several orders of magnitude greater than standard analysis pipelines.

For the BOSS analysis, the agreement between \effort{} and \pybird{} is similarly outstanding, as shown in Fig.~\ref{fig:boss}. We employed the same priors as in~\cite{DAmico:2022osl}, also including the correlated prior among EFT parameters. The \pybird{} chains, using the \montepython{} sampler, required a few days on a computing cluster to obtain chains with a very high degree of convergence. Using \effort{}, we ran eight chains for both NUTS and MCHMC, with the same burn-in and accepted steps as in the PT challenge. Chains were again initialized from a \texttt{Pathfinder.jl} run. The total wall-clock time was slightly over one hour, with ESS/s of 0.4 for NUTS and 2.4 for MCHMC.

\begin{figure}[ht!]
  \centering
  \includegraphics[width=0.9\textwidth]{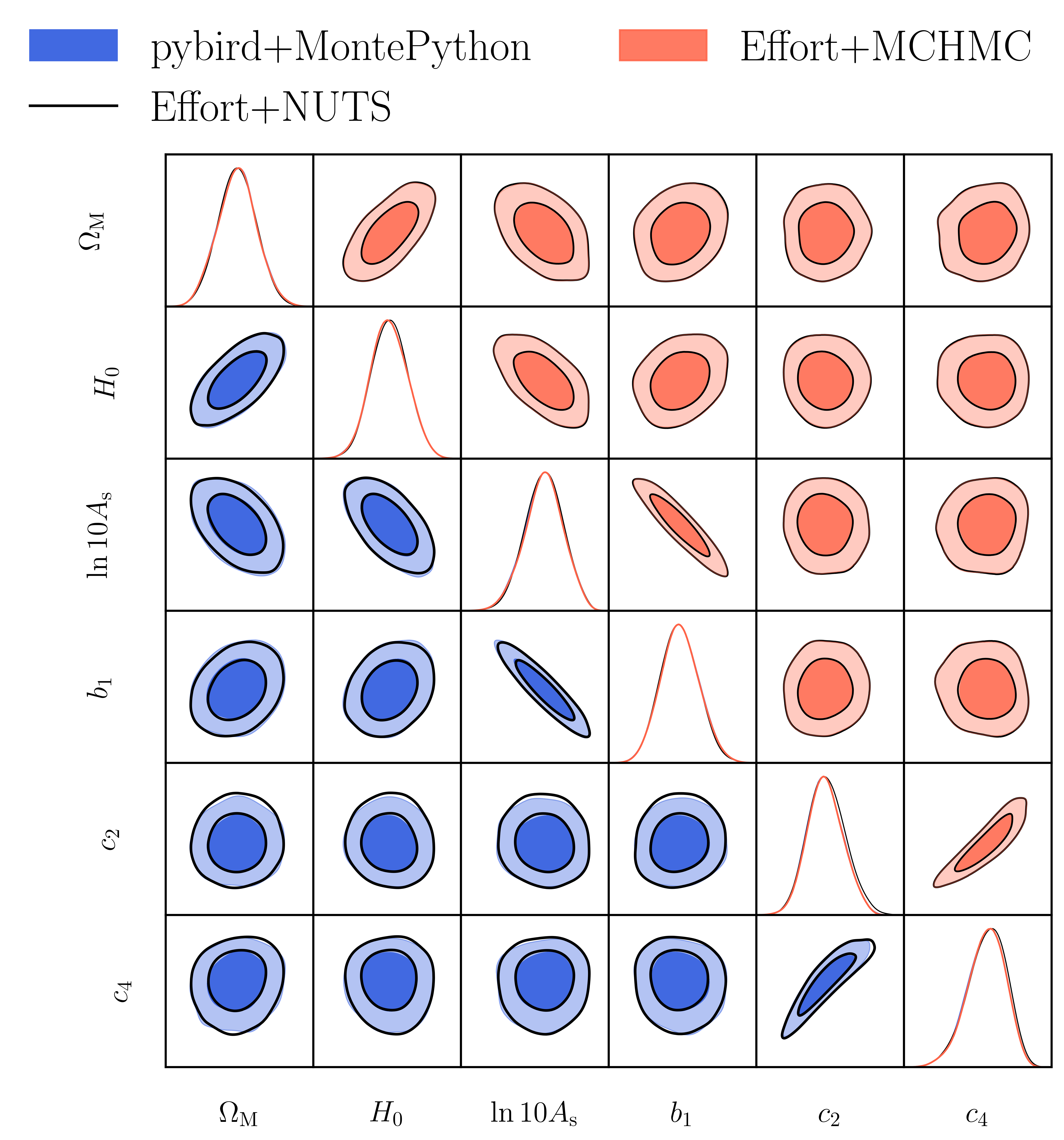} 
  \caption{Triangle plot showing the standard \pybird{} chains alongside those obtained using \effort{} in combination with \turing{}, for the PT-challenge analysis. The lower triangular section compares \pybird{} and \effort{} to validate the precision of our emulator and likelihood implementation. The upper triangular part focuses solely on \effort{} contours, comparing the results obtained with the NUTS and MCHMC samplers. To align with the PT-challenge guidelines, axis ticks and labels have been removed to avoid disclosing the true cosmology.}
  \label{fig:blinded_challenge}
\end{figure}

\begin{figure}[ht!]
  \centering
  \includegraphics[width=0.9\textwidth]{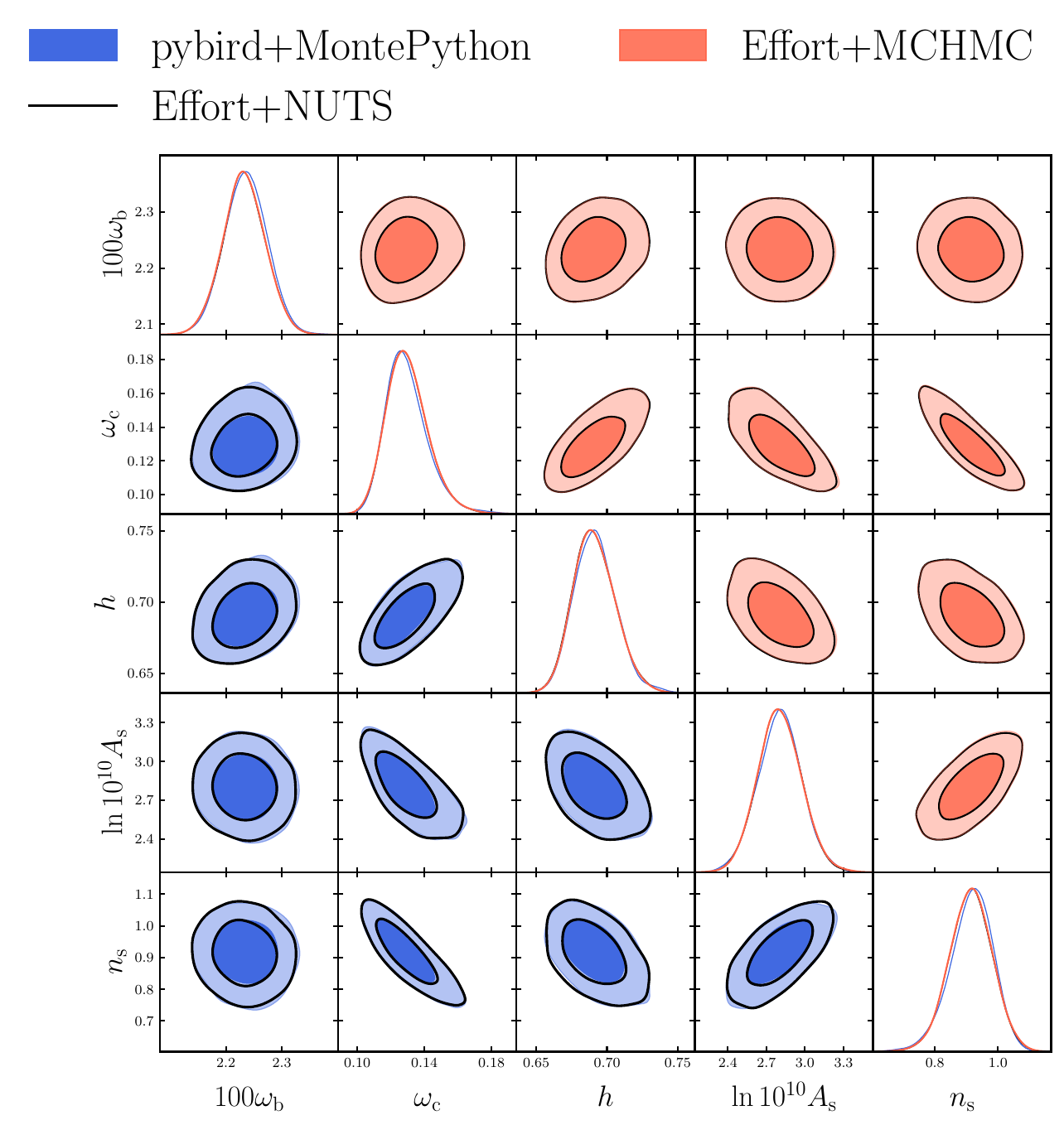} 
  \caption{Triangle plot showing the standard \pybird{} chains alongside those obtained using \effort{}, for the BOSS analysis, focusing on cosmological parameters. The lower triangular section compares \pybird{} and \effort{} to validate the precision of our emulator and likelihood implementation. As in fig.~\ref{fig:blinded_challenge}, in the lower triangular part we focus on the comparison with the \pybird{} chains, to ensure the correctness of our results, while in the upper part of the plot we compare the two gradient based samplers employed in this study.}
  \label{fig:boss}
\end{figure}

\section{Conclusions}
\label{sec:conclusions}
In this work, we introduced \effort{}, a novel emulator for the EFTofLSS with a strong emphasis on computational performance. The design of \effort{} prioritizes efficiency without compromising accuracy, leveraging state-of-the-art numerical methods and machine learning techniques to provide robust predictions for cosmological analyses.

Great care has been devoted to implementing observational effects, such as the Alcock-Paczynski effect and window mask convolution, ensuring high computational efficiency. This rigorous implementation enables \effort{} to handle the complexity of LSS data while maintaining precise modelling of observational effects. We checked the accuracy of these calculations both against standard Boltzmann solvers and \pybird{}, finding an excellent agreement.

We explored various preprocessing strategies, showcasing their significant impact on the emulator accuracy. Notably, we combined numerical method, such as solving ODE and symbolic regression, with NN emulators of cosmological observables. This hybrid approach demonstrates the potential for improving emulator precision and efficiency by leveraging domain knowledge readily available to cosmologists.

The coupling of \effort{} with the \turing{} framework highlights its versatility, allowing it to efficiently integrate with gradient-based samplers such as NUTS and MCHMC. Our results demonstrate the ability of \effort{} to sample Bayesian posteriors very efficiently, reducing computational costs significantly while achieving excellent agreement with standard pipelines. In particular, this will become of paramount importance when considering scenarios, like those found in~\cite{Ivanov:2024hgq,Paradiso:2024yqh, Akitsu:2024lyt, Shiferaw:2024ehr}, where analytical marginalization is not a viable option and we need to explore the full parameter space.

Looking forward, we plan to apply \effort{} to the forthcoming datasets from the DESI survey~\cite{DESI:2024hhd}, harnessing its computational efficiency to analyze next-generation cosmological data with precision. In addition, we foresee significant potential in combining \effort{} with tools such as \capse{}~\cite{capse}, which emulates CMB observables, and \blast{}~\cite{chiarenzablast}, which computes the photometric $3\times 2\,\mathrm{pt}$. This integration will facilitate efficient joint analyses of diverse cosmological datasets and probes. 

The combination of \effort{} and \blast{} is particularly compelling, as it enables the joint analysis of the Euclid main probes without relying on the Limber approximation, which can introduce inaccuracies in cosmological inference~\cite{1953ApJ...117..134L, Fang:2019xat}. Furthermore, as demonstrated in~\cite{piras2023}, the use of HMC samplers is expected to play a pivotal role in the analysis of $3\times 2\,\mathrm{pt}$ datasets from Stage IV surveys like Euclid. This approach will become even more critical when combining Euclid's photometric and spectroscopic datasets for joint analyses.

The capabilities of \effort{} provide a solid foundation for further enhancements and expansions. Additionally, we plan to extend the use of symbolic regression to $w_0w_a$ cosmologies, providing a computationally inexpensive replacement for the ODE computations, further optimizing the preprocessing pipeline.

The modular structure of \effort{} is sufficiently general to support compatibility with other EFT-based codes, such as \texttt{velocileptors}~\cite{Chen:2020zjt}, \texttt{CLASS-PT}~\cite{Chudaykin:2020aoj}, \texttt{FOLPS}~\cite{Noriega:2022nhf}, and \texttt{CLASS-OneLoop}~\cite{Linde:2024uzr}. This flexibility opens the door for training \effort{} to emulate these codes as well, broadening its application and usability.

Finally, we are actively working on a \texttt{jax}-based version of \effort{}. Since the cosmological community is more proficient with \texttt{python}, we thus see advantageous to provide a version of our software compatible with such a language. While a fully equivalent \texttt{jax} implementation of \capse{} is already complete\footnote{\href{https://github.com/CosmologicalEmulators/jaxcapse}{https://github.com/CosmologicalEmulators/jaxcapse}}, we are now focussing on translating \effort{} into \texttt{jax}.

\section*{Acknowledgements}
MB is supported in part by funding from the Government of Canada’s
New Frontiers in Research Fund (NFRF). MB also acknowledges the support of
the Canadian Space Agency and the Natural Sciences and Engineering Research Council of
Canada (NSERC), [funding reference number RGPIN-2019-03908].
This research was enabled in part by support provided by Compute Ontario (computeontario.ca) and the Digital Research Alliance of Canada (alliancecan.ca). MB acknowledges financial support from INAF MiniGrant 2022.
MB is expecially grateful to Anton Baleato-Lizancos, Guadalupe Canas-Herrera, Pedro Carrilho, Emanuele Castorina, Arnaud de Mattia, Minas Karamanis, Chiara Moretti,  Andrea Pezzotta,  Uros Seljak, and Martin White  for useful discussions. GDA is especially grateful to Pierre Zhang for useful discussions. We thank Takahiro Nishimichi for the PT-challenge simulations. Part of this work has been carried on the High Performance Computing facility of the University of Parma, whose support team we thank. MB thanks University of California Berkeley for hospitality during his visit, during which this project was started.

\appendix
\section{The galaxy power spectrum in the EFTofLSS}
\label{sec:eftoflss}
The Effective Field Theory of Large-Scale Structure (EFTofLSS) provides a systematic approach to modeling the redshift-space galaxy power spectrum by thoroughly accounting for the effects of small-scale physics on large-scale clustering. This framework extends standard perturbation theory by introducing counterterms that describe how small-scale physics, including galaxy formation, influences the observed large-scale distribution. We give a concise overview below and refer readers to \cite{Perko2016BiasedStructure,dAmico2020TheStructure} for more details.

The one-loop EFTofLSS expression for the redshift-space galaxy power spectrum is:
\begin{align}\label{eqn:gpk}
 P_{g}(k, \mu) & = Z_1(\mu)^2 P_{11}(k) 
 + 2 \int \frac{d^3q}{(2\pi)^3}\; Z_2(\mathbf{q},\mathbf{k}-\mathbf{q},\mu)^2 P_{11}(|\mathbf{k}-\mathbf{q}|)P_{11}(q)\nonumber  \\
& + 6 Z_1(\mu) P_{11}(k) \int\, \frac{d^3 q}{(2\pi)^3}\; Z_3(\mathbf{q},-\mathbf{q},\mathbf{k},\mu) P_{11}(q)\nonumber \\
& + 2 Z_1(\mu) P_{11}(k)\left( c_\text{ct}\frac{k^2}{{ k^2_\textsc{m}}} + c_{r,1}\mu^2 \frac{k^2}{k^2_\textsc{r}} + c_{r,2}\mu^4 \frac{k^2}{k^2_\textsc{r}} \right)\nonumber \\
& + \frac{1}{\bar{n}_g}\left( c_{\epsilon,0}+c_{\epsilon,1}\frac{k^2}{k_\textsc{m}^2} + c_{\epsilon,2} f\mu^2 \frac{k^2}{k_\textsc{m}^2} \right) \,,
\end{align}
where we follow the notation of \cite{DAmico2024TamingData}. This includes linear contributions, one-loop Standard Perturbation Theory (SPT) terms, counterterms, and stochastic terms. Here, $\mu$ is the cosine of the angle between the line of sight (LoS) and the wavenumber $\mathbf{k}$; $P_{11}(k)$ is the linear matter power spectrum; and $f$ is the growth factor. The scale $k_\textsc{m}^{-1}$ characterizes the size of collapsed objects, while $k_\textsc{r}^{-1}$ governs counterterms needed to handle products of the velocity field at the same point~\cite{DAmico2024TamingData,Ivanov2022CosmologicalDistortions}, and $\bar{n}_g$ denotes the mean galaxy number density.

The functions $Z_n$ are the redshift-space galaxy density kernels of order $n$:
\begin{align}\label{eqn:zkernels}
    Z_1(\mathbf{q}_1) & = K_1(\mathbf{q}_1) + f\mu_1^2 G_1(\mathbf{q}_1) = b_1 + f\mu_1^2\,, \nonumber\\ 
    Z_2(\mathbf{q}_1,\mathbf{q}_2,\mu) & = K_2(\mathbf{q}_1,\mathbf{q}_2) + f\mu_{12}^2 G_2(\mathbf{q}_1,\mathbf{q}_2) + \frac{1}{2}f \mu q \left( \frac{\mu_2}{q_2}G_1(\mathbf{q}_2) Z_1(\mathbf{q}_1) + \text{perm.} \right)\,, \nonumber\\ 
    Z_3(\mathbf{q}_1,\mathbf{q}_2,\mathbf{q}_3,\mu) & = K_3(\mathbf{q}_1,\mathbf{q}_2,\mathbf{q}_3) + f\mu_{123}^2 G_3(\mathbf{q}_1,\mathbf{q}_2,\mathbf{q}_3) \nonumber \\ 
    &\quad + \frac{1}{3}f\mu q \left(\frac{\mu_3}{q_3} G_1(\mathbf{q}_3) Z_2(\mathbf{q}_1,\mathbf{q}_2,\mu_{123}) +\frac{\mu_{23}}{q_{23}}G_2(\mathbf{q}_2,\mathbf{q}_3)Z_1(\mathbf{q}_1)+ \text{cyc.}\right)\,,
\end{align}
and $K_n$ are the corresponding $n$th-order galaxy density kernels in real space:
\begin{align}\label{eqn:kkernels}
    K_1 & = b_1\,, \nonumber\\
    K_2(\mathbf{q}_1,\mathbf{q}_2) & = b_1\frac{\mathbf{q}_1\cdot \mathbf{q}_2}{q_1^2} + b_2\left( F_2(\mathbf{q}_1,\mathbf{q}_2)- \frac{\mathbf{q}_1\cdot \mathbf{q}_2}{q_1^2} \right) + b_4 + \text{perm.} \,, \nonumber\\
    K_3(k, q) & = \frac{b_1}{504 k^3 q^3}\left( -38 k^5q + 48 k^3 q^3 - 18 kq^5 + 9 (k^2-q^2)^3\log \left[\frac{k-q}{k+q}\right] \right) \nonumber\\
    &\quad + \frac{b_3}{756 k^3 q^5} \left( 2kq(k^2+q^2)(3k^4-14k^2q^2+3q^4)+3(k^2-q^2)^4 \log \left[\frac{k-q}{k+q}\right] \right) \,.
\end{align}
We omit the full expressions for the second-order kernel $F_2$ and velocity kernels $G_n$ (from SPT) for brevity. Altogether, the model uses four bias parameters $b_1, b_2, b_3, b_4$, three counterterm parameters $c_{\rm{ct}}, c_{r,1}, c_{r,2}$, and three stochastic parameters $c_{\epsilon,0}, c_{\epsilon,1}, c_{\epsilon,2}$, for a total of ten parameters:
\begin{align}
    \{ b_1, b_2, b_3, b_4, c_{\rm{ct}}, c_{r,1}, c_{r,2}, c_{\epsilon,0}, c_{\epsilon,1}, c_{\epsilon,2} \}\,.
\end{align}
Finally, we employ IR-resummation \cite{Senatore2014RedshiftStructures, Senatore:2014via, Lewandowski2020AnPeak} to account for the large effects of long-wavelength displacements.

\bibliography{biblio}
\bibliographystyle{JHEP}

\end{document}